# Distribution of the Extinction and Star Formation in NGC 1569

M. Relaño[1], U. Lisenfeld[1,2], J.M. Vilchez[2], and E. Battaner[1]

[1] Dpto. Física Teórica y del Cosmos, Universidad de Granada, Spain
  e-mail: `mrelano@ugr.es,ute@ugr.es,battaner@ugr.es`
[2] Instituto de Astrofísica de Andalucía, CSIC, Apartado 3004, 18080, Granada, Spain
  e-mail: `jvm@iaa.es`

**Abstract.** We investigate spatial distribution of the intrinsic extinction in the starburst dwarf galaxy NGC 1569 creating an extinction map of the whole galaxy derived from the Hα/Hβ emission line ratio. We differentiate the extinction in the H ii regions from the extinction of the diffuse gas. The intrinsic extinction shows considerable variations over the plane of the galaxy, from negligible extinction up to highest values of $A_V = 0.8$ mag. The extinction map shows small scale clumpy structures possibly due to a clumpy dust distribution. We also identify in this map a shell structure, for which we establish a causal relation with the expanding gas structure produced by the stellar winds coming from the Super Star Clusters (SSC) in the center of the galaxy. The comparison of the spatial profiles of the extinction, dust and gaseous emissions crossing the border of the shell shows a layered structure; the peak of this Hα distribution lying closest to the SSC A, followed outwards by the peak of the extinction and at a still larger distances by the bulk of the atomic gas. We suggest that the extinction shell has been produced by the SSCs and that it can be explained by the accumulation of dust at the border of this ionized gas structure.

## 1. Introduction

Interstellar dust is an ubiquitous and important component of the interstellar medium. It can be studied through its emission or through the extinction it causes. In order to learn something about the dust properties and its distribution, good observations of the total galactic Spectral Energy Distribution (SED) are necessary (e.g. Popescu et al. 2000, Misiriotis et al. 2001; Tuffs et al. 2004) to constrain the dust heating and emission. The detailed observation of the dust emission has been difficult so far because of the poor transmission of the atmosphere. The heating of the dust is mainly due to the UV emission for which data has been lacking as well. Observations from GALEX and SPITZER are presently producing improvements (see, for example, Rowan-Robinson et al. 2005), which will soon be completed with the European spatial missions HERSCHEL and PLANCK.

The extinction towards extragalactic sites of star formation is usually being obtained from integrated measurements of the emitted fluxes coming from the H ii regions associated to these zones. The most extended way of obtaining the extinction of the light coming from an H ii region is based on the comparison of the Hα and Hβ recombination line fluxes (Caplan & Deharveng 1985, 1986; Maíz-Apellániz, Pérez & Mas-Hesse 2004), other estimates include radio–continuum measurements (e.g. Churchwell & Goss 1999; Viallefond, Goss

& Allen 1982) and wavelenghts in the Far–Ultraviolet (FUV) and Far–Infrared (FIR) ranges (e.g. Bell et al. 2002).

Comparison of the observations in the FIR/UV/Hα in a sample of H ii regions in the LMC shows that, although there is a spatial correlation between the emission in the FIR and Hα, implying a relation between the presence of the dust and the sites of star formation, the FIR and Hα emissions do not correlate spatially with the UV emission (Bell et al. 2002). These authors suggest that as the ionizing stars age (i.e. implying they do not emit more ionizing flux but they can still emit a significant fraction of UV flux), the distribution of the dust can decouple from the distribution of the ionizing stars due to the dispersal of the dust by the cumulative effects of stellar winds and supernovae. Thus, only young, ionizing stars are found to be spatially correlated with the dust emission whereas more evolved stars (though still emitting in the UV) are not.

The observed Hα extinctions estimated from the Balmer decrement are usually lower than the extinctions estimated from the comparison of the Hα and radio–continuum fluxes (Caplan & Deharveng 1986; van der Hulst et al. 1988). The effect is generally being attributed to inhomogeneities in the interior of H ii regions, which bias the extinction derived from the Balmer decrement towards lower values. The density inhomogeneities within the H ii regions have been modelled by Giammanco et al. (2004). However, observational evidence to explain the differences in extinction via inhomogeneities in the interstellar medium is still missing and observations with higher resolution are necessary to treat this issue.

Although it is tradionally known that dwarf galaxies have low metallicities and small amounts of dust, the distribution of the dust content within them can be strongly influenced by the star formation activity occurring in their interiors: (i) around sites of recent strong star formation there will be high UV-ionizing radiation fields that modify the dust properties in the surrroundings; and (ii) the evolved stars produce stellar winds and supernovae which interact with the interstellar gas and dust, modifying their distribution within the galaxies.

We investigate this issue in the nearby dwarf galaxy NGC 1569 by studying the extinction distribution over the whole disk of the galaxy. NGC 1569 is located at a distance of 2.2 Mpc (Israel 1988), which allows an adequate spatial resolution to study the structure of the interstellar gas and dust in its interior. Although it has a low Galactic latitude, which implies significant contamination by foreground Galactic extinction, its violent star formation and starburst phase makes it a very good object to study the relation of the emission and extinction of the dust with star formation. Values for local extinctions have been derived previously, revealing that the internal extinction within the galaxy is generally not very high. Devost, Roy & Drissen (1997) found a mean intrinsic value of $A_V$=0.65 mag, and Gonzalez-Delgado et al. (1997) obtained, close to the center of the galaxy, $A_V$=0.37 mag. Similar values were found by Origlia et al. (2001), $A_V$=0.47 mag, and by Kobulnicky & Skillman (1997), who found values of $A_V$ from 0.0 to 0.8 mag at different positions. In spite of the numerous studies which give a value for the internal extinction within NGC 1569, in none of them a map of the extinction for the whole galaxy has been obtained. The star formation history of NGC 1569 has been studied by e.g. Greggio et al. (1998), who showed that this galaxy experienced a global burst of star formation (SF) of $\geq$0.1 Gyr duration which ended ~5-10 Myr ago. During this period the Super Star Clusters (SSC) A and B were formed (see Hunter et al. 2000). Gonzalez-Delgado et al. (1997) present spectroscopic observations that suggest sequential star formation within these SSCs. The lack of ionized gas around the SSCs was ascribed by these authors to strong stellar winds and supernova explosions of the older burst, which expelled the gas from the vicinity of the clusters. Using long-slit spectroscopy, Martin (1998) identified an expanding shell of ionized gas centered at the central Star Clusters A and B of the galaxy.

In this paper we present high resolution H$\alpha$ and H$\beta$ images of NGC 1569 from the *Hubble Space Telescope* data archive. The good quality of the data allows us to study the H$\alpha$ extinction from the Balmer decrement at a very high resolution and produce a complete map of the extinction for the whole disk in this galaxy. In order to further analyze the extinction distribution, we have compared it to HI data, observations at 850 $\mu$m and CO data. Our aim here is to study the internal variations of the extinction in the whole dwarf galaxy and relate them to the strong effects (stellar winds and supernovae) of the star formation events.

## 2. Data from the Hubble Space Telescope

### 2.1. Data Reduction

NGC 1569 was observed in the H$\alpha$ and H$\beta$ emission lines on the 23rd of September of 1999. The WFPC2 camera in the *Hubble Space Telescope* (HST) observed NGC 1569 with three filters; two narrow ones, F656N and F487N, with effective filter widths of 28.3Å and 33.9Å, respectively, and a wider filter, F547M, with a width of 483.2Å. The first two were used to isolate the H$\alpha$ and H$\beta$ emission of the galaxy, while the third one was used for the continuum subtraction of the Balmer emission lines. The plate scale of the instrument is 0.09″/pix, which corresponds to 0.96 pc/pix at a distance of the galaxy of 2.2 Mpc. Since the H$\alpha$ and H$\beta$ emission line redshifts are ~2Å, they were observed close to the center of the transmission filters. The observations for the F656N filter consist of 4 exposures of 400 s, for the F487N filter of 4 exposures of 800 s, and for F547M of 2 exposures of 30 s. The data reduction of the observations in H$\beta$ is the same as that for H$\alpha$, thus we will explain here only the reduction for one of these observations. We started with the calibrated data from the HST archive which already have the bias and flatfield subtraction.

The four F656N exposures were aligned and combined, then cosmic rays were rejected using the STSDAS package *crrej* of the IRAF program. The WFPC2 mosaics were constructed using the task *wmosaic*. The same procedure was performed for the two exposures with the F547M filter and resulted in two image mosaics for the F656N and F547M observations. Using the keyword ORIENTAT in the image header we orientated the images in the North-South direction.

The H$\alpha$ on-line (F656N) and continuum (F547M) images were aligned using positions of field stars in both images and the scaling factor for the continuum subtraction was obtained using aperture photometry of these stars. The mean ratio of the continuum/on-line fluxes is 0.88, with a standard deviation of 0.17. We finally chose a scaling factor of 0.84 which gives the minimal residuals below 1$\sigma$ of the background emission after the continuum subtraction. A change in the scale factor of 0.17 produces an uncertainty in the H$\alpha$ flux of the H II regions less than 2%. We estimate a contribution of the nitrogen emission line [N II]$\lambda$6548 through the filter profile F656N expected to be less than 1% in the H$\alpha$ measured flux. The same procedure was followed for the H$\beta$ observations The scale factor obtained is 1.27$\pm$0.10, and we finally chose a scale factor of 1.09. Changes in the scale factor of 0.1 produce an uncertainty in the H$\beta$ flux of less than 6%.

### 2.2. Calibration

The On-The-Fly reprocessing system implemented by STScI ensures a calibration with the best–available reference files. Thus, the header keyword PHOTFLAM, defined as the mean flux density that produces a count rate of 1 s$^{-1}$, from the archive images should have the most appropiate value. For this study, we checked the PHOTFLAM keyword of the narrowband images, obtaining a synthetic value from the SYNPHOT package, which reproduces the photometric parameters of an indi-vidual observation. The task *bandpar* in the SYNPHOT package give the modified PHOTFLAM keyword (URESP) and the effective filter width. In Table 1 we show the values of the PHOTFLAM keywords of the retrieved images and the modified values, URESP. Differences between both values are minimal and we finally used URESP for our calculations.

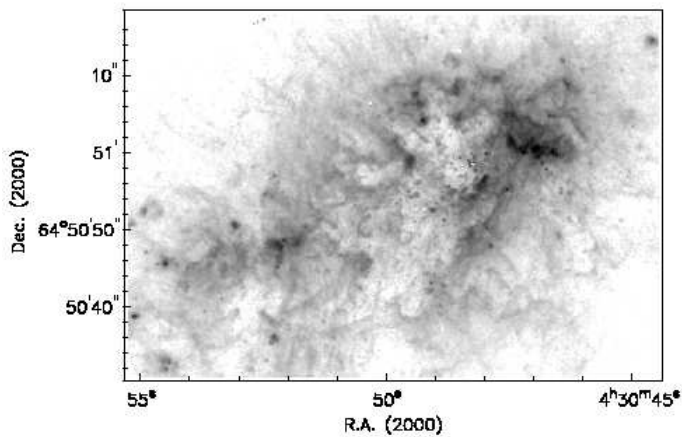

**Fig. 1.** Continuum–subtracted Hα emission of the galaxy NGC 1569. The pixel scale is 0.09″/pix and the *Point Spread Function* of the field stars has a FWHM of 0.3″.

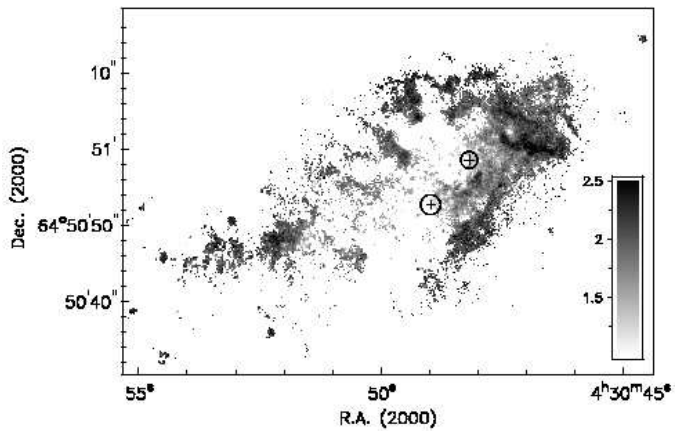

**Fig. 2.** Map of the logarithmic Hβ equivalent width of NGC 1569. The bar shows the greyscale for the values of the EW in Å. The circled crosses show the location of SSC A and B.

In order to check our calibration we compared our results with the integrated Hα flux of the H ɪɪ regions of NGC 1569 catalogued by Waller (1991). Using his estimated radius for each H ɪɪ region (see Table 3 of Waller 1991), we integrated the emission of the continuum-subtracted Hα image within the corresponding radius. The total count number for each H ɪɪ region is then multiplied by $1.473{\times}10^{-16}$counts$^{-1}$ erg s$^{-1}$ cm$^{-2}$Å$^{-1}$(URESP) and 28.3Å (filter width), and divided by the total exposure time, 1600 s. The Hα fluxes for the most luminous H ɪɪ regions coincide to better than 90% with the values given by Waller (1991) and the total Hα flux integrated within the Holmberg radius of the galaxy (1.4′, from de Vaucouleurs et al. (1991); RC3) is $2.07{\times}10^{-11}$ erg s$^{-1}$ cm$^{-2}$ s$^{-1}$, close to the value obtained by Waller, $2.04{\times}10^{-11}$ erg s$^{-1}$ cm$^{-2}$ s$^{-1}$. In Fig. 1 we show the continuum-subtracted Hα emission of the galaxy NGC 1569.

## 3. Extinction Maps

### 3.1. Influence of the stellar absorption

Before studying the extinction in the whole disc of NGC 1569 we have to take into account the influence of the underlying stellar absorption. This phenomenon can locally affect the calculation of the extinction. While for young hot stars, as those which are located within the H ɪɪ regions, the influence of the stellar absorption is expected to be small, for more evolved stars it can be important. In order to asses the problem we have created a map of the Hβ Equivalent Width (EW) within the whole face of NGC 1569 which is shown in Fig. 2. We find values of EW~25-50 Å at the position and in the vicinity of the SSCs A and B. In the rest of the galaxy we find values higher than 125Å and up to ~250Å in the center of the H ɪɪ regions. These values agree with those obtained previously by Kobulnicky & Skillman (1997) using spectrophotometry at several locations across NGC 1569 and by Devost et al. (1997), who presented spectroscopic data located at the positions of the most luminous H ɪɪ regions.

In order to estimate the extinction correction by the stellar absorption we refer to the value of EW=1-2Å derived by Kobulnicky & Skillman (1997) in NGC 1569. We have estimated the correction due to the stellar absorption in our measured Hβ flux using three measurements for the EW of 1Å, 1.5Å and 2Å, all of them in the range used by Kobulnicky & Skillman (1997) and including the estimate of Gonzalez-Delgado et al. (1997) who derived a stellar correction of 1Å for NGC 1569. We then generated maps of the Hβ flux corrected by 1Å, 1.5Å and 2Å EW.

In the bulk of the diffuse ionized gas (see section 3.2.1) the extinction correction for the underlying stellar absorption in the worst case of an EW correction of 2Å, can be up to ~-0.06 mag. Within the H ɪɪ regions (see section 3.2.2) and for the same EW correction, we obtain Hβ flux corrections of a factor of ~1.02, yielding an upper limit for the correction in the Balmer extinction of only ~-0.05 mag. Should we have included in addition the correction for the measured Hα flux, the final correction in the Balmer extinction would be less than this estimates. The EW due to the stellar absorption can be significative higher than 2Å (see table 5 of Gonzalez-Delgado, Leitherer & Heckman 1999) for an stellar population older than 4Myr. Using the values of the EW given in Fig. 2 and models of Starburst99 (Leitherer et al. 1999; for an instantaneous starburst, Salpeter Initial Mass Function and Z=0.004), we obtain that the stars located at the position of the H ɪɪ regions are younger than 3-4 Myr. Thus, the value of EW=2Å is a reasonable upper limit for the EW due to the underlying stellar absorption. The extinction corrections at detailed positions within the face of galaxy are further discussed in section 5.1.

**Table 1.** Calibration keyword values of the Hα and Hβ narrowband images.

| Filter | Emission Line | PHOTFLAM[a] | URESP[a] | Filter Width[b] | Total Exposure (s) |
|--------|---------------|-------------|----------|-----------------|--------------------|
| F487N  | Hβ            | 3.964×10⁻¹⁶ | 3.966×10⁻¹⁶ | 33.9          | 3200               |
| F656N  | Hα            | 1.473×10⁻¹⁶ | 1.473×10⁻¹⁶ | 28.3          | 1600               |

[a] Units of counts⁻¹ erg s⁻¹ cm⁻² Å⁻¹.

[b] Units of Å.

At the positions of the SSCs A and B, giving their age, $\tau \geq 7$ Myr (e.g. Hunter et al. 2000, Gonzalez-Delgado et al. 1997), which implies the presence of evolved stars, the Balmer decrement is so strongly affected by stellar absorption that a correction is difficult. The extinction there can best be derived by spectroscopy. Kobulnicky & Skillman (1997) did this and found values of $A_{H\alpha} = 1.3$ at the position of the SSC A and B. This corresponds to the value of the Galactic extinction assumed here (see section 3.4) and shows that the intrinsic extinction at the position of the SSCs is negligible.

### 3.2. Method to calculate the extinction

The procedure to obtain the Hα extinction from the Hα and Hβ emission line fluxes is well known and explained e.g. in Caplan & Deharveng (1986). These authors assume that the Balmer lines are produced in thermodynamical equilibrium conditions and that departures from the theoretically expected value for the Hα/Hβ ratio is produced by extinction of the light due to a screen of homogeneously distributed interstellar dust. Assuming a standard extinction curve ($R_V=3.1$[1]), these authors derived the following expression for the extinction of Hα produced by such a dust screen distribution:

$$A_{H\alpha}(\text{Balmer}) = 5.25 \times \log_{10}\left(\frac{H\alpha/H\beta}{2.859 t_e^{-0.07}}\right) \quad (1)$$

where Hα/Hβ is the ratio of the observed Hα and Hβ emission line fluxes, $t_e$ the electronic temperature in units of $10^4$ K, and 2.85 represents the expected Hα/Hβ ratio for case B recombination at T=$10^4$ K and electron density of $100\,\text{cm}^{-3}$(Brocklehurst 1971).

Using the high resolution Hα and Hβ HST images, we obtain a map of the Hα/Hβ line ratio of NGC 1569. We first eliminated in both images (noisy) pixels that had an intensity level below three times the r.m.s.. The result is shown in Fig.3. We see that the Hα/Hβ line ratio is above 2.85 practically everywhere across the galaxy, indicating the presence of extinction. Furthermore, very small-scale variations of the ratio are noticeable in this figure and will be further discussed in Sect. 5.2.

Since here we are interested in the distribution of the extinction at the global scale of the galaxy, we have smoothed our high-resolution (∼0.3″) HST images to a lower resolution of 6″, corresponding to ∼65pc in NGC 1569. This is the scale of the kinematic features observed by Martin (1998) within the galaxy and a good compromise to improve our signal-to-noise but still keeping information at a reasonable spatial resolution. In Fig. 4 we show the smoothed (6″) Hβ image with smoothed Hα intensity contours overlaid. The two most intense peaks correspond to H ii regions 2 and 7 in Waller (1991). Apart from the emission coming from H ii regions, we can see wide-spread diffuse emission, as previously detected by Waller (1991). In the following, we will discuss the extinction in these two different regimes separately, because of two reasons: (i) The physical conditions can differ for the two gaseous components and (ii) the diffuse emission might be partly contaminated by shock-ionization in contrast to the emission in H ii regions, which is produced by photoionization.

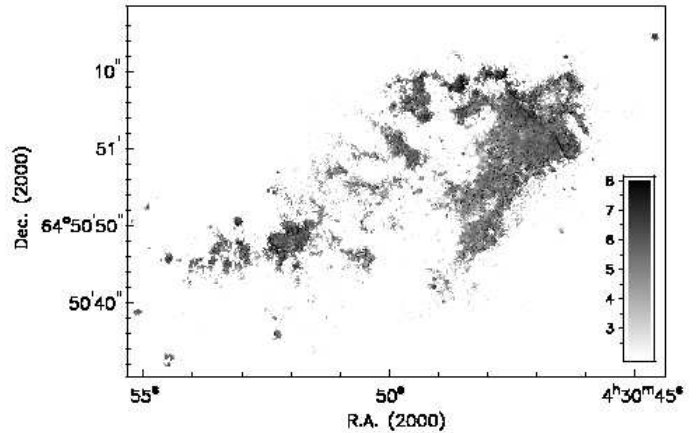

**Fig. 3.** High resolution (0.3″) Hα/Hβ line ratio map obtained from the continuum-subtracted Hα and Hβ images. The bar shows the grayscale for the values of the Hα/Hβ line ratio.

### 3.2.1. Derivation of the extinction of the diffuse ionized gas

The diffuse Hα emission is known to be more extended and with a lower surface brightness than the emission coming from H ii regions. In order to separate the two components, we estimate the limit of the diffuse ionized emission of the galaxy from the frequency distributions of the Hα emission at several positions. We explain here the results obtained from the high resolution data (Fig. 1); for the smoothed (6″) Hα emission map an analogous procedure was followed. In Fig. 5 we show the frequency distribution of the Hα surface brightness for a zone of the galaxy which includes an H ii region and an

---

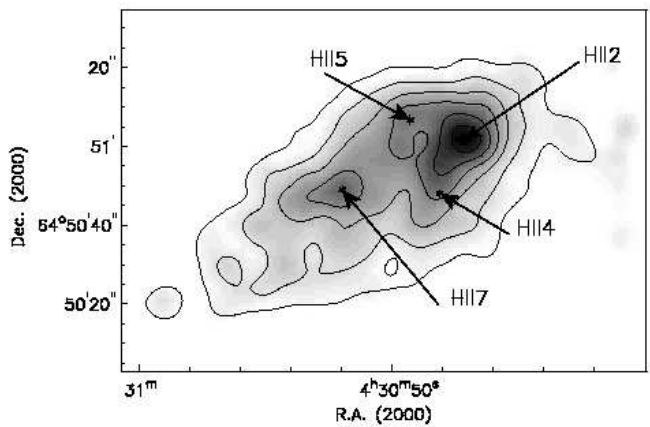

**Fig. 4.** Smoothed Hβ image at 6″ resolution, overlaid with contours of the Hα emission.

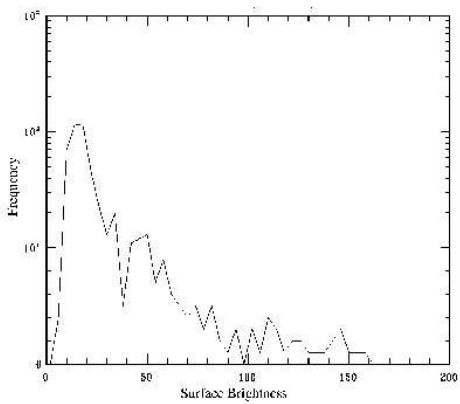

**Fig. 5.** Frequency distribution of the Hα emission in a box of the high resolution Hα image (Fig. 1), which includes both an isolated H ɪɪ region and a more extended diffuse emission. The surface brightness units are in counts/pixels. The points with very high surface brightness and low frequency correspond to the emission of the H ɪɪ region, while the points with low surface brightness and high frequency correspond to the diffuse emission outside the H ɪɪ region.

extended zone outside the H ɪɪ region. In this figure we can separate two components: the diffuse ionized emission is characterized by a low surface brightness with high frequencies which are decreasing rapidly towards higher surface brightness, whereas the emission of the H ɪɪ regions has a high surface brightness but with lower frequencies and a flat distribution. From Fig. 5 we adopt 80 counts/pixel, which corresponds to an Hα flux of $2.1 \times 10^{-16}$ erg s$^{-1}$ cm$^{-2}$, as an estimate for the transition value between both regimes and as an upper limit for the diffuse ionized emission. As a test, we integrated the emission of the galaxy below 80 counts per pixel within the size of the galaxy (1.4′, from RC3) and obtained a flux of $5.2 \times 10^{-12}$ erg s$^{-1}$ cm$^{-2}$, which corresponds to 25% of the total Hα flux of NGC 1569. This fraction coincides with the limits to the diffuse ionized gas emission of low Hα luminosity spiral galaxies obtained by Zurita et al. (2000).

After isolating the emission below $2.1 \times 10^{-16}$ erg s$^{-1}$ cm$^{-2}$ we can create a map for the Hα extinction in the diffuse gas using Eq. 1 and adopting appropriate physical conditions for this regime. Reynolds & Cox (1992) give a typical temperature of 6000 K for the warm ionized gas taking into account grain heating. A study of the Local diffuse gas by Haffner et al. (1999) showed that the warm ionized medium in the Galaxy can have temperatures in the range of 6000 to 10000 K, increasing from faint to bright Hα emission. Following these studies, we will adopt a temperature for the diffuse ionized gas (DIG) of 6000 K. A change of 2000 K in the temperature would imply an error of 0.05 in $A_{H\alpha}$.

A further source of error in the derivation of the extinction is the possible shock-excitation of part of the diffuse ionized gas in addition to the photo-ionization which Eq. 1 relies. Long–slit spectra observations of 14 dwarf galaxies including NGC 1569 carried out by Martin et. al (1997), suggest that the dominant excitation mechanism of the DIG is photoionization, although a significant contribution (up to 30%-50% of the emission) from shock-excited gas is also noted. With this value as a guide and the Hα/Hβ line ratio given by Dopita & Sutherland (1996) of 3.2 for a shock velocity of 150 km s$^{-1}$ we can estimate the error produced. Assuming a contamination from shock-excited gas of 40% in the Hα/Hβ line ratio (Martin et al. 1997), we estimate a difference in extinction of $-0.1$ mag with respect to pure recombination. For lower velocity shocks of $\sim 50$ km s$^{-1}$, we extrapolate Table 8 of Dopita & Sutherland (1996) and derive a value of the Hα/Hβ ratio of $\sim3.4$; this gives a difference of $-0.15$ mag in $A_{H\alpha}$. Thus, in the worst case, assuming pure photoionization, we could be overestimating $A_{H\alpha}$ by 0.15 magnitudes in sites where shocks are present.

### 3.2.2. Derivation of the extinction towards the H ɪɪ regions

H ɪɪ regions are defined in our images, as described in the last section, as those regions where the surface brightness lies above $1.0 \times 10^{-14}$ ( erg s$^{-1}$ cm$^{-2}$) for the 6" smoothed maps. Within these H ɪɪ regions, we derived the extinction assuming a constant electron temperature in Eq. 1. Following Kobulnicky & Skillman (1997), who found an electronic temperature of T(O$^{++}$)$\sim$12400 K for several H ɪɪ regions in NGC 1569, and assuming a typical difference of $\sim$-1500 K for T(H$^+$), we will consider an electron temperature of 11000 K for our H ɪɪ regions. A change of the adopted temperature by 1500 K would produce a difference in the Balmer extinction of 0.02 magnitudes.

### 3.3. Extinction maps

The total extinction map $A_{H\alpha}$ of the diffuse emission is shown in Fig.6. In this image we have excluded the diffuse outer halo emission of NGC 1569, with a very low surface brightness because of (i) the uncertainty in its temperature, which can be different from the one of the diffuse interstellar medium in the disk and (ii) it would require a study of the outer galaxy halo, which is out of the scope of this paper. We have derived the limits in surface brightness for the halo from the mean Hα and Hβ emission in the western arm of the galaxy, plus three times the rms of the emission in this zone. In the extinction map in Fig.6 an internal structure is clearly visible with typical values between 1.0 mag and 1.4 mag.

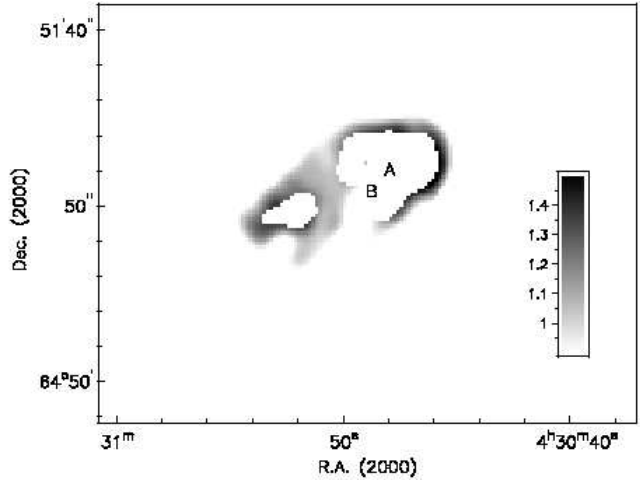

**Fig. 6.** Extinction map of the diffuse Hα emission in NGC 1569 obtained from the ratio of the Hα and Hβ emission fluxes. The resolution of the image is 6″. The letters refer to the position of the Star Clusters A and B, and the bar shows the grayscale for the values of the extinction $A_{Hα}$ in magnitudes.

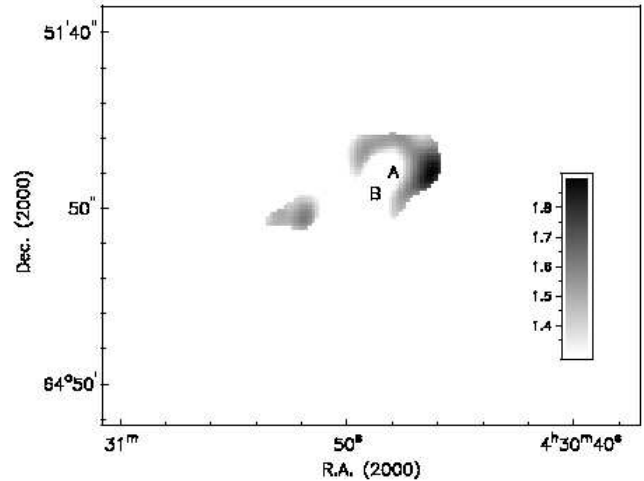

**Fig. 7.** Extinction map (6″) for the H II regions in NGC 1569 obtained from the ratio of the Hα and Hβ emission fluxes. The letters refer to the position of the Star Clusters A and B and the bar has the same meaning as in Fig. 6.

In Fig. 7 we show the extinction for the H II regions. The most prominent visible structure is the arc formed by the highest values of extinction, which resembles a shell located near the position of H II region 2 of Waller (see Fig. 4). A further extinction maximum coincides with H II region 7 of Waller. The maximum extinction within the shell structure is located towards the western side of H II region 2. This shell structure matches very well with complex C found by Martin (1998) expanding from the center of the positions of the Star Clusters A and B. As we will see later, the expanding shell affects the extinction values derived for this galaxy. The minima in Fig. 7 match the positions of the Star Clusters A and B, coinciding with the hole in HI detected by Israel & Driel (1990). We have not estimated the extinction at the position the two SSCs because of the effect of the corresponding stellar absorption features of the clusters, and the very low surface brightness of the ionized gas there (see section 3.1)..

Finally, in Fig. 8 we show the extinction map as a combination of the extincion maps for H II regions and the diffuse gas, overlaid with the extinction contours. The shell in extinction is clearly seen for extinction values higher than ∼1.5 magnitudes. Nevertheless, the exact range adopted for the extinction variation within the galaxy in Figs. 6, 7, and 8 must take into account the correction due to the underlying stellar absorption, as discussed in section 3.1.

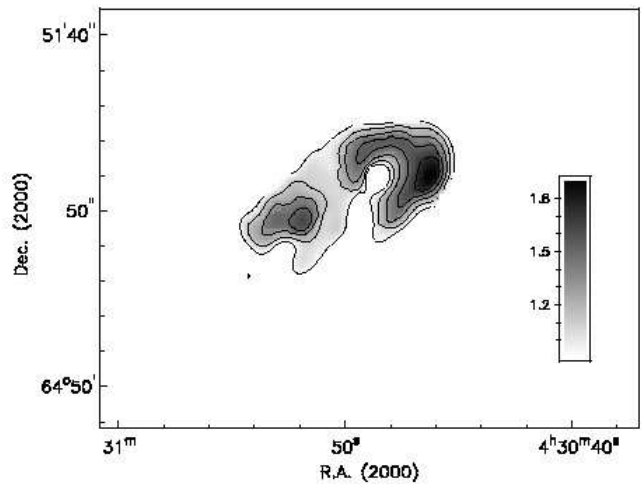

**Fig. 8.** Extinction map (6″) for the galaxy NGC 1569 obtained as a composition of the extinction for the diffuse gas and the H II regions. The extinction contours are at $A_{Hα}$ =0.9, 1.1, 1.3, 1.5, 1.7 magnitudes and the bar shows the adopted grayscale. The extinction values could be overestimated up to a maximum of 0.05 mag due to the underlying stellar absorption (see section 3.1). For the maximum uncertainties due to changes of the physical conditions of the ionized gas see sections 3.2.1 and 3.2.2.

### 3.4. Galactic and intrinsic extinction towards NGC 1569

Due to the relatively low galactic latitude of NGC 1569, part of the extinction we have derived comes from the Galaxy. The study of the Galactic contribution to the extinction has been carried out by several authors. Following Burstein & Heiles (1984) the foreground extinction for NGC 1569 is $A_B$=2.03, which corresponds to $A_V$=1.53 (following the extinction curve of table 4 in Draine 2003). Schlegel, Finkbeiner & Davis (1998) give higher Galactic extinction, $A_V$(Gal) =2.32. On the other hand, Israel (1988) found a value for the Galactic reddening of E(B-V)=0.55, which gives, adopting R=3.1, $A_V$=1.71.

Devost et al. (1997) studied the extinction across the face of NGC 1569 and found $A_V = 1.61 \pm 0.09$ due to our own Galaxy and a mean intrinsic extinction of $< A_V >_{\text{intrinsic}}= 0.65 \pm 0.04$. He obtained values for the total $A_V$ between 1.9 and 2.6 magnitudes (see his Fig. 5), which corresponds (following the extinction curve of table 4 in Draine 2003) to values of $A_{H\alpha}$ of 1.6 and 2.0 magnitudes, similar to the values obtained in this paper (see Fig. 7). Gonzalez-Delgado et al. (1997) found an intrinsic extinction of $A_V$=0.37 at about 8″ to the northwest of SSC A and a total extinction at this position of $A_V$=2.08, which corresponds to $A_{H\alpha}$=1.62, similar to the value found here at this position (see Fig. 8). Origlia et al. (2001) obtained a Galactic contribution to the extinction of $A_V$=1.71 ($A_{H\alpha}$=1.33) and an intrinsic extinction of $A_V$=0.47 ($A_{H\alpha}$=0.37) close to the SSC A. Variations of the intrinsic extinction in NGC 1569 were found by Kobulnicky & Skillman (1997) with long-slit optical spectrophotometry. They found values for $A_V$ from 0.0 to ∼0.80 magnitudes ($A_{H\alpha}$ ∼0.6) showing that the intrinsic extinction in NGC 1569 is clearly not spatially constant. Both, the values and the spatial distribution of the intrinsic extinction found by them agree with our results.

We will adopt here a value of the Galactic extinction of $A_V = 1.64$ ($A_{H\alpha} = 1.28$), a mean from the $A_V$ values provided by Israel (1988), Origlia et al. (2001), Devost et al. (1997) and Burstein & Heiles (1984). We do not include here the Galactic extinction given by Schlegel et al. (1998), since it is above the values derived for the *total* extinction by most of the other authors. Therefore, we think that it must be an overestimate for the Galactic extinction at the position of NGC 1569. As a conservative estimate of the uncertainty of this mean value we adopt the difference between the mean and the extreme values considered, yielding 0.09 mag. The value for the Galactic extinction assumed here coincides well with the lower limit of the extinction derived by us. The only places where our extinction is below this limit is (i) at the position of SSC A and B and (ii) at places of very low S/N of $H\alpha$ and $H\beta$ emission. As stated above, the extinction derived at the position of the star clusters is not reliable because of the substantial stellar absorption, yielding an unreliable $H\alpha/H\beta$ ratio. We assume here the extinction values obtained by Kobulnicky & Skillman (1997), $A_{H\alpha} = 1.3$ at the position of the SSC A and B, which shows that the intrinsic extinction at the position of the SSCs is negligible and much lower than in the shell region surrounding them. Even though we have excluded in our analysis the positions in the map with a very low surface brightness in $H\alpha$ and $H\beta$ the error in the regions with the lowest emission is still relatively large (∼0.3mag). Taking into account this error, the value for the extinction found here agrees with the lower limit of the Galactic extinction.

Once subtracted the Galactic emission we found that in many regions of the diffuse emission there is non-measurable intrinsic extinction. The maximum values obtained for the intrinsic Balmer extinction within the H II regions is 0.6 mag. Also Kobulnicky & Skillman (1997) found a pronounced in-crease of the extinction along the slit towards the North and West of the SSC A and B, consistent with the position of our shell shown in Fig. 8. This shows that although globally NGC 1569 has a low average intrinsic extinction, the internal extinction variation, especially that related to the shell structure, is substantial and it will be discussed in section 5. As we have pointed out before, extinction laws different from the one used here have been derived for other starburst galaxies, and this fact would produce a change of the absolute values of the extinction. Although this is an important effect, it will not change the conclusions of this paper devoted to the analysis of the relative extinction variations in the galaxy.

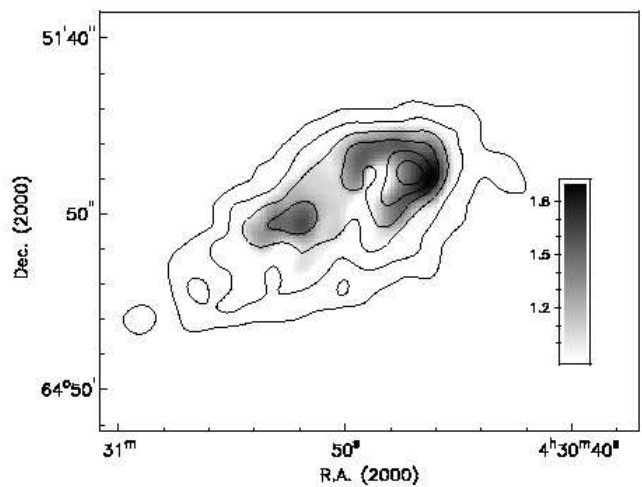

**Fig. 9.** Composed extinction map (6″) obtained from the ratio of the $H\alpha$ and $H\beta$ emission fluxes with contours overlaid of the $H\alpha$ emission at a resolution of 6″. The bar has the same meaning as in previous figures and the extinction uncertainties are given in Fig. 8.

## 4. Extinction structure and the distributions of the ionized, neutral and molecular gas and dust

In Fig. 9 we show the extinction map of the whole galaxy (combination of the maps for H II regions and diffuse gas) overlaid with contours of the $H\alpha$ emission. Although there is a general resemblance between $H\alpha$ and the extinction maps, the latter shows a much clearer shell structure and the maximum of the extinction lies westwards of the $H\alpha$ maximum of the most luminous H II region 2.

In order to compare the distribution of dust emission and extinction, we show in Fig. 10 the extinction, $A_{H\alpha}$, overlaid with contours of the 850 $\mu$m dust emission taken from Lisenfeld et al. (2002) at a resolution of 15″. A certain correlation between the global distribution of the extinction and dust emission (see e.g. the contour level at 35 mJy/beam) is noticeable. There seems to be a slight offset of ∼ 3″ between the maximum of the 850 $\mu$m emission and the maximum of the dust emission. However, the low resolution of the 850 $\mu$m data does not allow us to ascertain whether it is real or not.

The positions of the Giant Molecular Clouds (GMC) found by Taylor et al. (1999) from observations of the $^{12}$CO 2→1 at 2″ resolution, though located close to H II region 2 (see Fig. 10) do not coincide exactly with the position of the highest values of the extinction in this region. The molecular cloud closest to the peak in extinction, and coinciding with the peak of the 850 μm emission, is GMC 3 of Taylor, which is the least luminous of the three clouds.

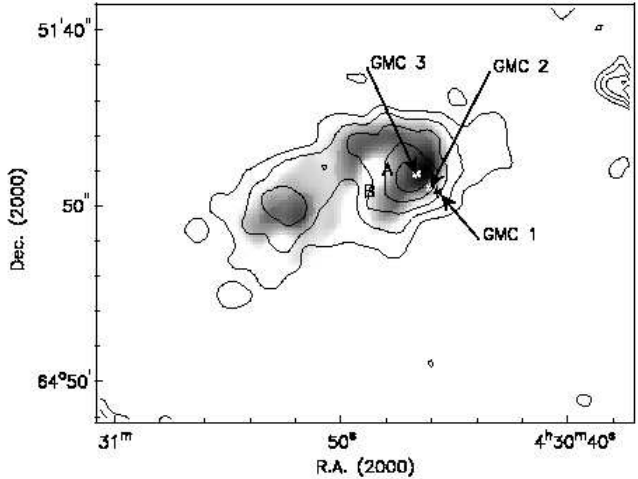

**Fig. 10.** Hα extinction map (6″ resolution) with contours of the dust emission at 850μm at a resolution of 15″. Contour values are at levels 15, 25, 35, 45 and 55 mJy/beam. The asterisks mark the position of the Giant Molecular Clouds detected by Taylor et al. (1999). The uncertainties in extinction as given in Fig. 8.

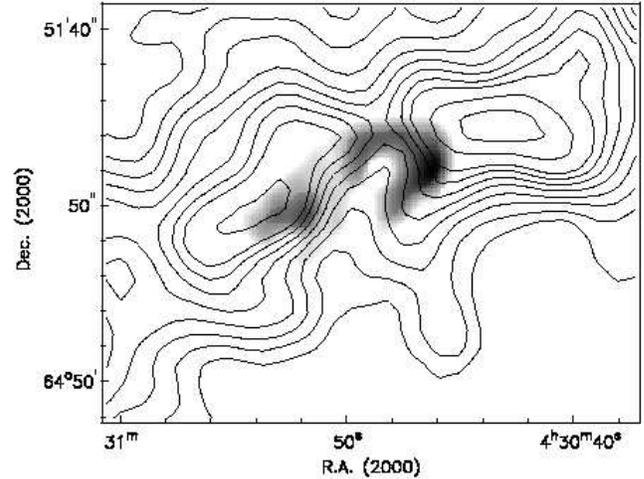

**Fig. 11.** Extinction map ($A_{Hα}$) with contours overlaid of HI intensity from Stil & Israel (2002) at 13.5″. The isointensity levels are in $N(HI) = (1.5, 2.0, 2.5, 3.0, 3.5, 4.0, 4.5, 5.0, 6.0, 7.0) \times 10^{21}\,cm^{-2}$. The uncertainties in extinction as given in Fig. 8.

Finally, we have also checked whether there is a relation between the internal extinction structure of NGC 1569 and its HI distribution. We have overlaid the contours of the HI intensity from Stil & Israel (2002) at 13.5″ resolution on the extinction map in Fig. 11. The HI contours mark very well the lowest $A_{Hα}$ extinction structure surrounding the HI hole at the position of the SSCs. Furthermore, an HI ridge surrounds the western and southern part of the extinction shell.

## 5. Discussion

### 5.1. Origin of the distribution of the extinction

In Fig. 8, a *shell structure* marked by the highest values of the extinction in the north, west and south-westside of the galaxy can be clearly seen. The shell covers the location of HII regions 2, 4 and 5 of Waller (1991) and is centered around the position of the Star Cluster A, with an average radius of 90 pc. It corresponds to the supershell NGC 1569-C catalogued by Martin (1998) (see her Fig. 2k), which is centered at the Star Cluster A and has a radius of 85 pc and an expansion velocity of 79 kms$^{-1}$.

We suggest that this shell structure has been formed by an accumulative deposit of dust in the boundary of the supershell found by Martin (1998). To test this scenario, we first check whether this shell could have been produced by stellar winds coming from the Star Cluster A. Adopting a radius of 90 pc and a velocity of 79 kms$^{-1}$ for the shell, following Martin (1998), we obtain a kinematical age for the supershell of $\sim 10^6$ yr, which is thus consistent within the age of the Star

Cluster A, $\tau \geq 7$ Myr, derived by Hunter et al. (2000). Assuming a SFR of 0.5 M$_\odot$yr$^{-1}$, a Salpeter IMF and the Geneva tracks with Z=0.001, proposed by Greggio et al. (1998) to explain the stellar population of NGC 1569, we used the models from Starburst 99 (Leitherer et al. 1999) to estimate the wind kinetic energy produced by Star Cluster A. Taking the theoretical value of 0.2 given by Dyson (1980) for the efficiency of the stellar wind to convert its energy into interstellar kinetic energy, we obtained that the Star Cluster A has injected an energy of $E_K(ISM) = 3.2 \times 10^{52}$erg into the interstellar medium. This energy is higher than the kinetic energy of the shell C, estimated by Martin (1998) to be $E_K(C) = 1.4 \times 10^{52}$erg. The energy of the stellar winds can also account for the kinetic energy of the dust in the shell. Taking the total dust mass of NGC 1569 from the model of Lisenfeld et al. (2002), ($M_{dust} \sim 1.0 \times 10^6 M_\odot$) and assuming, based on the distribution of the dust emission in Fig. 10 that about 50% of the dust is situated in the shell region, we estimate the kinetic energy necessary to move this mass to ∼80 kms$^{-1}$ to be $E_K(dust) = 3.0 \times 10^{51}$erg, which is an order of magnitude smaller than the kinetic energy injected by the Star Cluster A.

We have only taken into account the kinetic energy deposited from the Star Cluster A, which is high enough to move the shell, but radiative energy, typically two orders of magnitude higher, can also contribute energetically to the formation and expansion of the shell. Martin (1998) has also found that the star formation in NGC 1569 could power the largest expanding complexes. Based on these arguments we conclude, that energetically and kinematically the scenario in which SSC A has caused the observed shell structure is plausible.

In Fig. 12 we show the spatial profiles of the extinction $A_{Hα}$, Hα intensity, dust emission at 850 μm and HI column density. The profiles were obtained starting at the position of SSC A and going eastwards at PA=90°, thereby crossing the shell at its maximum in extinction (see Fig. 9). We also include an esti-mate of the error in extinction at different positions of the profile due to the uncertainties in the temperature of the gas in the H II regions and the effect for the correction by a stellar absorption of EW=1.5Å. The figure reveals a spatial displacement of the maxima of the different profiles: closest to the position of SSC A is the peak of the Hα emission (indicating the center of H II region 2), followed by the maximum in $A_{H\alpha}$ which is displaced from the Hα peak by ∼ 6″ to the west, whereas the maximum of the 850 µm emission lies somewhere in between. The low resolution (15″) makes it impossible to ascertain whether there is a closer relation of the 850 µm profile to the dust extinction or to the Hα emission. However, Lisenfeld et al. (2002) found a good coincidence of the peak of the dust emission at 1.2mm (with a better spatial resolution of 10″), with the position of H II region 2, supporting a close relation between the cold dust and Hα emission. Outside the extinction shell is the peak of the HI distribution, which traces the HI ridge seen in Fig. 11.

We interpret this distribution as a consequence of dust accumulating in the outer region of the ionized shell and surrounding a region where ionized gas and dust could be partially mixed. In the more central parts of the H II region the dust can be mixed with the gas producing relatively little extinction, especially if its distribution is clumpy (see the next section). Therefore, a line of sight in this direction, crossing only a small layer of the dust-rich shell border, would show less extinction than a line of sight along the border of the shell.

## 5.2. Small–scale extinction structure in the H II regions and the diffuse ionized gas

At the location of H II region 2, the Hα/Hβ emission line ratio map shown in Fig. 3 has a very clumpy structure with knots reaching Hα/Hβ ∼ 6 − 7 located close to zones with values of Hα/Hβ around 4 − 5. These fluctuations cannot be explained only by noise: in the interior of H II region 2 the calculated relative error of the Hα/Hβ quotient is between 0.1-0.2, whereas the measured disperion in the emission line ratio within the same region can reach 0.5-0.6. Thus, we have fluctuations in the Hα/Hβ ratio that are typically 3-6 times higher than the calculate noise in the map.

In order to quantify the frequency of these fluctuations we have excluded in H II region 2 all pixels with values of Hα/Hβ less than 5.0 (corresponding to the Galactic Hα extinction of 1.3 mag). We find that in an box covering the whole H II region 2 approximately 30% of the pixels have Hα/Hβ ratios larger than 6.0, corresponding to an *intrinsic* Hα extinction of 0.4 magnitudes.

We suggest that these fluctuations in the Hα/Hβ ratio can be related to opacity inhomogeneities in the interior of the H II regions, most likely due to a clumpy dust distribution. Similar inhomogeneities in the gas distribution have already been modelled realistically by Giammanco et al. (2004), suggesting that such a clumpy dust distribution could also be present within the H II regions here. We can estimate a limit to the size of these dust clumps from the spatial scale of the variations observed in the extinction map. The typical distance between maximum and minimum extinction measured in our map is about 1″, which gives a typical linear scale of the dust inhomogeneities of about 10 pc.

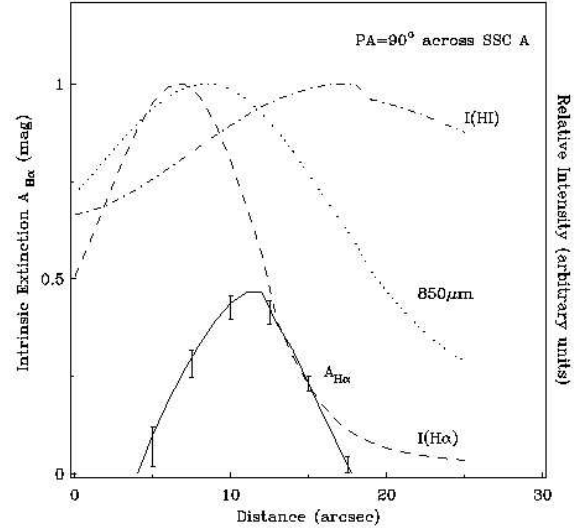

**Fig. 12.** Spatial profiles starting at the position of the SSC A and going eastwards at a position angle of 90°, crossing the maximum of the extinction shell. The profiles of I(Hα), I(HI) and 850 µm are shown normalized to their maxima (in arbitrary units) to compare them with the extinction profile. $A_{H\alpha}$ refers to the intrinsic extinction within NGC 1569 after subtracting the Galactic contribution. The error bars in the extinction profile are the combination of the errors in the temperature estimate (0.02 mag, see section 3.2.2) and the effect of the stellar absorption correction by EW=1.5Å: -0.08, -0.05, -0.04, -0.04, -0.02, -0.02 magnitudes at the positions 5, 7.5, 10, 12.5, 15, and 17.5 arcseconds from SSC A, respectively.

## 5.3. Variations of extinction per unit gas mass

We have compared the value of the extinction to the HI column density at various locations of NGC 1569. The variations of N(HI) over the region of interest in the galaxy are relatively small with values in the range of $(3 − 5) \times 10^{21}$cm$^{-2}$. The intrinsic (i.e. after subtraction of the Galactic foreground extinction) values of $A_{H\alpha}$ vary between non-detectable extinction and 0.6 mag (i.e. $A_V$ = 0.8 mag.). We estimate the uncertainty in this value to be roughly 0.1 mag, including uncertainties in the calculation of the extinction (0.02 mag, see section 3.2.2) and the Galactic foreground contribution (0.09 mag, see section 3.4). The highest value of the extinction occurs at the shell surrounding H II region 2 and the lowest values coincide with the HI hole south of SSC A and B. Adopting for the minimum intrinsic extinction an upper limit of 0.2 mag we derive values for the extinction per H atom between $A_{H\alpha}/N(HI) \leq 0.6 \times 10^{-22}$ mag cm$^2$ for the region of the HI hole and $A_{H\alpha}/N(HI) \approx (1.2 \pm 0.2) \times 10^{-22}$ mag cm$^2$ for the position of maximum extinction on the shell, corresponding to $A_V/N(HI) \approx (1.6 \pm 0.3) \times 10^{-22}$ mag cm$^2$. We conclude that the extinction per HI atom shows a variation of a factor $\sim 2$ over the disk of NGC 1569. The variation of the extinction per hydrogen particle (ie. including both atomic and molecular gas) might however vary much less. The maximum extinction at the shell is close to the position of molecular gas detected in CO (Greve et al. 1996). The molecular gas mass associated to this cloud is uncertain due to the poorly known conversion factor in low-metallicity environment. Adopting a Galactic conversion factor yields $N(H_2) = 3 \times 10^{20} cm^2$, a value more than 10 times lower than the HI gas surface density results. Lisenfeld et al. (2002) estimate however that the true conversion factor in NGC 1569 is much higher, giving a molecular column density of $N(H_2) = 8 \times 10^{21} cm^2$, slightly higher than the HI column density. In this case we would get $A_{H\alpha}/N(H + H_2) = (0.5 \pm 0.1) \times 10^{-22}$ mag $cm^2$, similar to the lowest values found for the region of the HI hole.

This range of values can be compared to the Galactic extinction per HI atom for which Bohlin (1978) derived $A_V/N(HI) = 8 \times 10^{-22}$ mag $cm^2$. The recent value by Fitzpatrick (1999) is slightly lower, $A_V/N(HI) = 6 \times 10^{-22}$ mag $cm^2$. These values translate to $A_{H\alpha}/N(HI) = 6.4 \times 10^{-22}$ mag $cm^2$, respectively $4.8 \times 10^{-22}$ mag $cm^2$. The highest values found in NGC 1569 are only a factor of $3 - 6$ below the Galactic value. The comparison to the total (molecular and atomic) gas column density increases the difference to a factor $8 - 16$.

## 6. Conclusions

We have studied the distribution of the extinction in the starbursting dwarf galaxy NGC 1569 from the Balmer $H\alpha$ and $H\beta$ emission lines. Our main conclusions from the analysis are:

- The intrinsic extinction considerable variations over the disk of the galaxy. The maximum value of the intrinsic extinction in $H\alpha$ is about $A_{H\alpha} = 0.6$ mag, corresponding to an extinction in the V-band of $A_V = 0.8$.
- The extinction in the North-West of the galaxy shows a pronounced shell structure around the SSC A. It coincides spatially with an expanding shell found by Martin et al. (1998). The shell shows a layered structure: the extinction derived from the Balmer decrement peaks outside the $H\alpha$ intensity maximum. The 850 $\mu$m dust emission, with a poorer resolution of 15 ″, has its maximum somewhere in between those emissions. Still further outside is the peak of the HI distribution, the shape of which outlines vaguely the shell structure of the extinction. We interpret this distribution as dust accumualated at the outer border of a shell produced by the SSCs A and B.
- We have studied the energetics of the shell and found that the energy from the stellar winds ejected by the SSC A is able to produce the shell. Kinematically, the size of the shell is also consistent with its origin from the SSC A.
- In the high-resolution $H\alpha/H\beta$ image we find substantial intrinsic small scale structure. We suggest that these arise from the variations of the dust opacity within the H II regions, due to a clumpy dust distribution.
- The highest values of the extinction per atomic hydrogen column density are a factor of about $3-6$ below the Galactic

value. This value could however go down to a factor of about $8 - 16$ when taking into account also the (uncertain) molecular gas column density.

*Acknowledgements.* This work has been supported by the Spanish Ministry of Education, via PNAYA (Spanish National Program for Astronomy and Astrophysics), the research projects AYA 2005-07516-C02-01, AYA 2004-08260-C03-C02 and ESP 2004-06870-C02-02, and the Junta de Andalucía. Thanks to Veronica Melo for her comments on the calibration process of the WFPC2 images. Thanks to Stefanie Mühle and Jeroen Stil for their HI data and to Jorge Iglesias Páramo for his useful comments. We thank the referee for constructive comments which helped us to make significant improvements to the article.

## References

Bell, E. F., Gordon, K. D., Kennicutt, J. R., Zaritsky, D. 2002, ApJ, 565, 994
Bohlin, R. C., Savage, B. D., Drake, J. F. 1978, ApJ, 224, 132
Brocklehurst, M. 1971, MNRAS, 153, 471
Burstein, D., Heiles, C. 1984, ApJSS, 54, 33
Calzetti, D. 2001, PASP, 113, 1449
Caplan, J., Deharveng, L. 1986, A&A, 155, 297
Caplan, J., Deharveng, L. 1985, A&ASS, 62, 63
Churchwell, E., Goss, W. M. 1999, ApJ, 514, 188
de Vaucouleurs G., de Vaucouleurs A., Corwin, H. G., et al. 1991, Third Reference Catalogue of Bright Galaxies, Springer, New York (RC3)
Devost, D., Roy, J. R., Drissen, L. 1997, ApJ, 482, 765
Dopita, M. A., Sutherland, R. S. 1996, ApJS, 102, 161
Draine, B. T. 2003, ARA&A, 41, 241
Dyson, J. E., 1980, Physics of the Interestellar Medium, (New York: John Wiley & Sons)
Fitzpatrick, E. L. 1999, PASP, 111, 755, 63
Giammanco, C., Beckman, J. E., Zurita, A., Relaño, M. 2004, 424, 877 755, 63
Gonzalez-Delgado, R. M., Leitherer, C., Heckman, T. M., Cerviño, M. 1997, ApJ, 483, 705
Gonzalez-Delgado, R. M., Leitherer, C., Heckman, T. M. 1999, ApJS, 125, 489
Leitherer, C., Schaerer, D., Goldader, J.D., Gonzalez-Delgado, R. M. et al. 1999, ApJS, 123, 3
Greggio, L., Tosi, M., Clampin, M., De Marchi, G., Leitherer, C., Nota, A., Sirianni, M. 1998, ApJ, 504,725
Greve, A., Becker, R., Johansson, L. E. B., Mc Keith, C. D. 1996, A&A, 312, 391
Haffner, L. M., Reynolds, R. J., Lufte, S. L. 1999, ApJ, 523, 233
Hunter, D. A., O'Connell, R. W., Gallagher, J. S., Smecker-Hane, T. A. 2000, ApJ, 120, 2383
Israel, F. P. 1988, A&A, 194, 24
Israel, F. P., Driel, W. v. 1990, A&A, 236, 323
Kobulnicky, H. A., Skillman, E. 1997, ApJ, 489, 636
Leitherer, C., Schaerer, D., Goldader, J. D. et al. 1999, ApJSS, 123, 3
Lisenfeld, U., Israel, F. P., Stil, J. M., Sievers, A. 2002, A&A, 382, 860
Maíz-Apellániz, J., Pérez, E., Mas-Hesse, J. M. 2004, AJ, 128, 1196
Misiriotis, A., Popescu, C. C., Tuffs, R., Kylafis, N. D., 2001, A& A, 372, 775
Martin, C. L. 1997, ApJ, 491, 561
Martin, C. L. 1998, ApJ, 506, 222
Origlia, L., Leitherer, C., Aloisi, A., Greggio, L., Tosi, M. 2001, ApJ, 122, 815

Popescu, C. C., Misiriotis, A., Kylafis, N. D., Tuffs, R. J., Fischera, J. 2000, 362, 138

Schlegel, D. J., Finkbeiner, D. P., Davis, M. 1998, ApJ, 500, 525

Reynolds, R. J., Cox, D. P. 1992, ApJ, 400, L33

Rowan-Robinson, M., Babbedge, T., Surace, J., Shupe, D., Fang, F., Lonsdale, C., Smith, G. et al. 2005, AJ, 129, 1183

Stil, J. M., Israel, F. P. 2002, A&A, 392, 473

Taylor, C. L., Hüttemeister, S., Klein, U., Greve, A. 1999, A&A, 349, 424

Tuffs, R. J., Popescu, C. C., Völk, H. J., Kylafis, N. D., Dopita, M. A. 2004, 419, 821

van der Hulst J. M., Kennicutt, R. C., Craine, P. C., Rots, A. H. 1988, A&A, 195, 38

Viallefond, F., Goss, W. M., Allen R. J. 1982, A&A, 115, 373

Waller, W. H. 1991, ApJ, 370, 144

Zurita, A., Rozas, M., Beckman, J. E. 2000, A&A, 363, 9